\documentclass{iaus}
\usepackage{graphicx}

\title[JD 11.~~The Milky Way chemodynamics] 
{The chemodynamical evolution of the Milky Way disc -- A new modeling approach}

\author[Ivan Minchev, Cristina Chiappini \& Marie Martig]   
{Ivan Minchev$^1$,
Cristina Chiappini$^1$
 \and Marie Martig$^2$}

\affiliation{$^1$Leibniz-Institut f\"{ur} Astrophysik Potsdam (AIP), \\ 
An der Sternwarte 16, D-14482, Potsdam, Germany \\
email: {\tt iminchev1@gmail.com} \\[\affilskip]

$^2$Centre for Astrophysics \& Supercomputing, Swinburne University of Technology, \\
P.O. Box 218, Hawthorn, VIC 3122, Australia}

\pubyear{2014}
\volume{298}  
\pagerange{XX--YY}
\jname{IAUS\,298 Setting the scence for Gaia and LAMOST}
\editors{S. Feltzing, G. Zhao, N.\,A. Walton \& P.\,A. Whitelock, eds.}
\begin{document}

\maketitle

\begin{abstract}
Despite the recent advancements in the field of galaxy formation and evolution, fully self-consistent simulations are still unable to make the detailed predictions necessary for the planned and ongoing large spectroscopic and photometry surveys of the Milky Way disc. These difficulties arise from the very uncertain nature of sub-grid physical energy feedback within models, affecting both star formation rates and chemical enrichment. To avoid these problems, we have introduced a new approach which consists of fusing disc chemical evolution models with compatible numerical simulations. We demonstrate the power of this method by showing that a range of observational results can be explained by our new model.  
We show that due to radial migration from mergers at high redshift and the central bar at later times, a sizable fraction of old metal-poor, high-[$\alpha$/Fe] stars can reach the solar vicinity. This naturally accounts for a number of recent observations related to both the thin and thick discs, despite the fact that we use thin-disc chemistry only. Within the framework of our model, the MW thick disc has emerged naturally from (i) stars born with high velocity dispersions at high redshift, (ii) stars migrating from the inner disc very early on due to strong merger activity, and (iii) further radial migration driven by the bar and spirals at later times. A significant fraction of old stars with thick-disc characteristics could have been born near the solar radius.

\end{abstract}

\firstsection 
\section{Introduction}

Crucial information regarding the dominant mechanisms responsible for the formation of the Milky Way (MW) disc and other Galactic components is encoded in the chemical and kinematical properties of its stars. This realization is manifested in the number of on-going and planned spectroscopic Milky Way surveys, such as RAVE (\cite[Steinmetz et al. 2006]{steinmetz06}; Boeche et al., this volume), SEGUE (\cite[Yanny et al. 2009]{yanny09}), 
APOGEE (\cite[Majewski et al. 2010]{majewski10}; Schultheis et al., this volume), 
HERMES (\cite[Freeman 2010]{freeman10}), 
Gaia-ESO (\cite[Gilmore et al. 2012]{gilmore12}), 
Gaia (\cite[Perryman et al. 2001]{perryman01}; Vallenari et al., this volume), 
LAMOST (\cite[Zhao et al. 2006]{zhao06}; Deng et al, this volume) and 
4MOST (\cite[de Jong et al. 2012]{dejong12}), which aim at obtaining kinematic and chemical information for a large number of stars. 

The common aim of these observational campaigns is to constrain the MW assembly history -- one of the main goals of the field of Galactic Archaeology. The underlying principle of Galactic Archaeology is that the chemical elements synthesized inside stars, and later ejected back into the interstellar medium (ISM), are incorporated into new generations of stars. As different elements are released to the ISM by stars of different masses and, therefore, on different timescales, stellar abundance ratios provide a cosmic clock, capable of eliciting the past history of star formation and gas accretion of a galaxy. In most cases, the stellar surface abundances reflect the composition of the interstellar medium at the time of their birth, which is the reason why stars can be seen as fossil records of the Galaxy evolution. One of the most widely used ``chemical-clocks" is the [$\alpha$/Fe] ratio\footnote{The notation in brackets indicates abundances relative to the Sun, i.e., [X/Y] $= \log$(X/Y)$ - \log$(X/Y)$_{\odot}$.}. 

In order to be able to interpret the large amounts of data to come, galaxy formation models tailored to the MW are needed.

\section{Difficulties with fully self-consistent simulations}
\label{sec:cos}

Producing disc-dominated galaxies has traditionally been challenging for cosmological models. In early simulations, extreme angular momentum loss during mergers gave birth to galaxies with overly-concentrated mass distributions and massive bulges (e.g., \cite[Navaro et al. 1991, Abadi et al. 2003]{navarro91,abadi03}). Although an increase in resolution and a better modeling of star formation and feedback have allowed recent simulations to produce MW-mass galaxies with reduced bulge fractions (\cite[Agertz et al. 2011, Guedes et al. 2011, Martig et al. 2012]{agertz11, guedes11, martig12}), none of these simulations include chemical evolution. Galaxy formation simulations including some treatment of chemical evolution have been performed by several groups (e.g., \cite[Raiter et al. 1996, Kawata et al. 2003, Scannapieco et al. 2005, Few et al. 2012]{raiteri96, kawata03, scannapieco05, few12}). 

\begin{figure}
\begin{center}
 \includegraphics[width=5.4in]{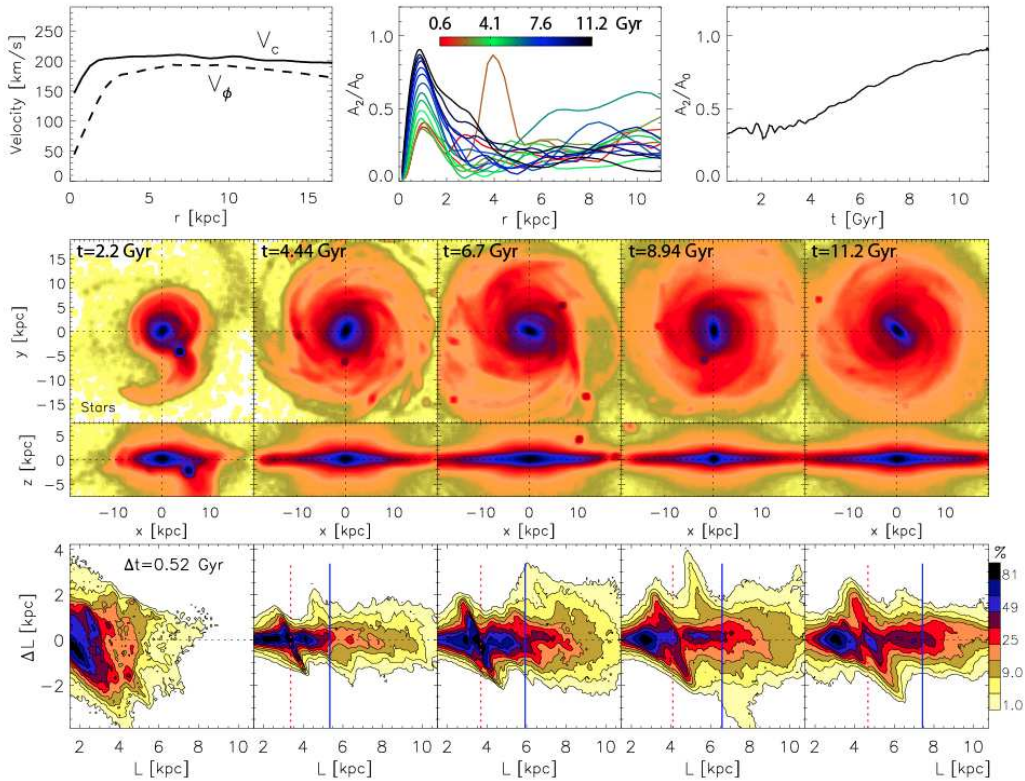} 
 \caption{
 {\bf First row:} The left panel shows the rotational velocity (dashed blue curve) and circular velocity (solid black curve) at the final simulation time. The middle panel presents the $m=2$ Fourier amplitudes, $A_2/A_0$, as a function of radius estimated from the stellar density. Curves of different colors present the time evolution of $A_2/A_0$. To see better the evolution of the bar strength with time, in the right panel we show the amplitude averaged over the bar maximum. {\bf Second row:} Face-on density maps of the stellar component for different times, as indicated. {\bf Third row:} The corresponding edge-on view. Contour spacing is logarithmic {\bf Fourth row:} Changes in angular momentum, $\Delta L$, as a function of radius, estimated in a time window of 0.52~Gyr, centered on the times of the snapshots shown above. Both axes are divided by the circular velocity, thus units are kpc (galactic radius). Strong variations are seen with cosmic time due to satellite perturbations and increase in bar strength. 
 }
   \label{fig1}
\end{center}
\end{figure}

However, although the results are encouraging, and global observed trends seem to be reproduced, such as the mass-metallicity relation Ê(e.g., \cite[Kobayashi et al. 2007]{kobayashi07}, Kobayashi et al., this volume) or the metallicity trends between the different galactic components (e.g., \cite[Tissera et al. 2012]{tissera12}), it is still a challenge for these simulations to reproduce the properties of the MW (e.g., the typical metallicities of the different components -- Ê\cite[Tissera et al. 2012]{tissera12}, Tissera et al., this volume). Additionally, the fraction of low metallicity stars are often overestimated (\cite[Kobayashi et al. 2011, Calura et al. 2012]{kobayashi11,calura12}), and reproducing the position of thin- and thick-disc stars in the [O/Fe]-[Fe/H] plane has proved challenging (\cite[Brook et al. 2012]{brook12}). While such issues could be due to unresolved metal mixing (\cite[Wiersma et al. 2009]{wiersma09}), it is also worth noting that none of the above-mentioned simulations reproduce simultaneously the mass, the morphology and star formation history (SFH) of the MW.

This situation has led us to look for a novel way to approach this complex problem. We will show that this approach works encouragingly well, explaining not only current observations, but also leading to a more clear picture regarding the nature of the MW thick disc (see also Bensby et al., this volume).

\section{A new chemo-dynamical model}
\label{sec:model}

To properly model the MW, it is crucial to be consistent with some observational constraints at redshift $z=0$, for example, a flat rotation curve, a small bulge, a central bar of an intermediate size, gas to total disc mass ratio of $\sim0.14$ at the solar vicinity, and local disc velocity dispersions close to the observed ones.

It is clear that cosmological simulations would be the natural framework for a state-of-the-art chemodynamical study of the MW. Unfortunately, as described in Sec.~\ref{sec:cos}, a number of star formation and chemical enrichment problems still exist in fully self-consistent simulations. We, therefore, resort to (in our view) the next best thing -- a high-resolution simulation in the cosmological context coupled with a pure chemical evolution model (see \cite[Minchev et al. 2013]{minchev12c}), as described below.

\subsection{A late-type disc galaxy simulation in the cosmological context}

The simulation used in this work is part of a suite of numerical experiments first presented by \cite{martig12}, where the authors studied the evolution of 33 simulated galaxies from $z=5$ to $z=0$ using the zoom-in technique described in \cite[Martig et al. (2009)]{martig09}. This technique consists of extracting merger and accretion histories (and geometry) for a given halo in a $\Lambda$-CDM cosmological simulation, and then re-simulating these histories at much higher resolution (150~pc spatial, and 10$^{4-5}$~M$_{\odot}$ mass resolution). The interested reader is referred to \cite[Martig et al. (2012)]{martig12} for more information on the simulation method. 

\begin{figure}
\includegraphics[width=5.3in]{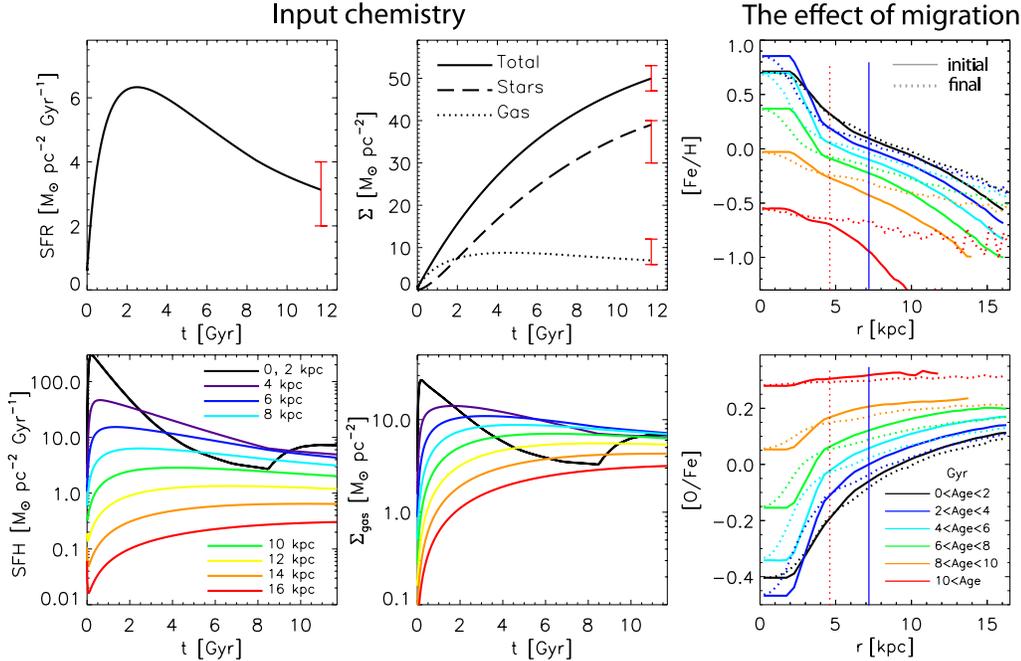}
\caption{
Properties of our detailed thin-disc chemical evolution model: 
{\bf Top left:} SFR in the solar neighborhood resulting from model. {\bf Top middle:} The total (solid), stellar (dashed) and gas (dotted) density time evolution at the solar neighborhood. The error bars are observational constraints (see \cite[Minchev et al. 2013]{minchev12c} for references to the data).
{\bf Bottom left:} SFR as a function of time for different galactic radii as color-coded. 
{\bf Bottom middle:} Total gas density (remaining after stars formation + recycled) as a function of time for different galactic radii as color-coded on the left. 
{\bf Top right:} [Fe/H] input (solid curves) and final (dotted curves) gradients. Different colors correspond to different look-back times as indicated. The dotted-red and solid-blue vertical lines indicate the positions of the bar's CR and OLR at the final simulation time.
{\bf Bottom right:} As on top but for the [O/Fe] gradients.
}
\label{fig2}      
\end{figure}

Originally, our galaxy has a rotational velocity at the solar radius of 210 km/s and a scale-length of $\sim5$~kpc. To match the MW in terms of dynamics, at the end of the simulation we downscale the disc radius by a factor of 1.67 and adjust the rotational velocity at the solar radius to be 220 km/s, which affects the mass of each particle according to the relation $GM\sim v^2r$, where $G$ is the gravitational constant. This places the bar's corotation resonance (CR) and 2:1 outer Lindblad resonance (OLR) at $\sim4.7$ and $\sim7.5$~kpc, respectively, consistent with a number of studies (e.g., \cite[Dehnen 2000]{dehnen00}, \cite[Minchev et al. 2007, 2010]{mnq07, minchev10}; Monari et al., this volume; Antoja et al., this volume). At the same time the disc scale-length, measured from particles of all ages in the range $3<r<15$~kpc, becomes $\sim3$~kpc, in close agreement with observations. After this change{\footnote{Note that we rescale the disc radius so as to reflect the position of the bar's resonances at the final time.}, our simulated disc satisfies the criteria outlined at the beginning of this Section, required for any dynamical study of the MW, in the following:

(i) it has an approximately flat rotation curve with a circular velocity $V_c\sim220$~km/s at 8~kpc, shown in the first row, left panel of Fig.~\ref{fig1}, where we have corrected for asymmetric drift as described. 

(ii) the bulge is relatively small, with a bulge-to-total ratio of $\sim$1/5 (as measured with GALFIT -- \cite[Peng et al. 2002]{peng02} -- on a mock i-band image, see bottom panel of Fig.~27 by \cite[Martig et al. 2012]{martig12}). 

(iii) it contains an intermediate size bar at the final simulation time, which develops early on and grows in strength during the disc evolution (see middle and right panels of Fig.~\ref{fig1} and discussion below).

(iv) the disc grows self-consistently as the result of cosmological gas accretion from filaments and (a small number of) early-on gas-rich mergers, as well as merger debris, with a last significant merger concluding $\sim9-8$~Gyr ago. 

(v) the disc gas-to-total mass ratio at the final time is $\sim$0.12, consistent with the estimate of $\sim0.14$ for the solar vicinity (Fig.~\ref{fig2}, top middle panel). 

(vi) the radial and vertical velocity dispersions at $r\approx8$~kpc are $\sim45$ and $\sim20$~km/s (averaged over all ages), in very good agreement with observations (see Fig.~6 in \cite[Minchev et al. 2013]{minchev12c}).

In the first row, middle panel of Fig.~\ref{fig1} we plot the Fourier amplitude, $A_m/A_0$, of the density of stars as a function of radius, where $A_0$ is the axisymmetric component and $m$ is the multiplicity of the pattern; here we only show the $m=2$ component, $A_2/A_0$. Different colors indicate the evolution of the $A_2$ radial profile in the time period specified by the color bar. The bar is seen to extend to $\sim3-4$~kpc, where deviations from zero beyond that radius are due to spiral structure. The brown curve reaching $A_2/A_0\sim0.9$ at $r\approx4.5$~kpc results from a merger-induced two-armed spiral. To see better the evolution of the bar strength with time, in the right panel we show the amplitude averaged over 1~kpc at the bar maximum. 

The second and third rows of Fig.~\ref{fig1} show face-on and edge-on stellar density contours at different times of the disc evolution, as indicated in each panel. The redistribution of stellar angular momentum, $L$, in the disc as a function of time is shown in the fourth row, where $\Delta L$ is the change in the specific angular momentum as a function of radius estimated in a time period $\Delta t=520$~Myr, centered on the time of each snapshot above. Both axes are divided by the rotational velocity at each radius, therefore $L$ is approximately equal to the initial radius (at the beginning of each time interval) and $\Delta L$ gives the distance by which guiding radii change during the time interval $\Delta t$. The dotted-red and solid-blue vertical lines indicate the positions of the bar's CR and OLR. Note that due to the bar's slowing down, these resonances are shifted outwards in the disc with time.

After the initial bulge formation, the largest merger has a 1:5 stellar-mass ratio and an initially prograde orbit, plunging through the center later and dissolving in $\sim1$~Gyr (in the time period $1.5\lesssim t\lesssim2.5$~Gyr, first column of Fig.~\ref{fig1}). Due to its in-plane orbit (inclination $\lesssim45^\circ$), this merger event results in accelerated disc growth by triggering strong spiral structure in the gas-dominated early disc, in addition to its tidal perturbation. One can see the drastic effect on the changes in angular momentum, $\Delta L$, in the fourth row, right panel of Fig.~\ref{fig1}. 
A number of less violent events are present at that early epoch, with their frequency decreasing with time. The effect of small satellites, occasionally penetrating the disc at later times, can be seen in the third and fourth columns at $L\approx r\gtrsim7$ and $L\approx r\approx6$~kpc, respectively. 

We note a strong variation of $\Delta L$ with cosmic time, where mergers dominate at earlier times (high $z$) and internal evolution takes over at $t=5-6$~Gyr (corresponding to a look-back-time of $\sim$6-7~Gyr, or $z\sim1$). The latter is related to an increase with time in the bar's length and major-to-minor axes ratio as seen in the face-on plots, indicating the strengthening of this structure. Examining the top right panel of Fig.~\ref{fig1}, we find a continuous increase in the bar's $m=2$ Fourier amplitude with time, where the strongest growth occurs between $t$=4 and 8~Gyr. The effect of the bar can be found in the changes in angular momentum, $\Delta L$, as the feature of negative slope, centered on the CR (dotted-red vertical line), shifting from $L\approx3.4$ to $L\approx4.7$~kpc. Due to the increase in the bar's amplitude, the changes in stellar guiding radii (vertical axis values) induced by its presence in the CR-region double in the time period  $4.44<t<11.2$~Gyr (bottom row of Fig.~\ref{fig1}). Until recently bars were not considered effective at disc mixing once they were formed, due to their long-lived nature. We emphasize the importance of the bar in its persistent mixing of the inner disc {\it throughout the galactic evolution} (see \cite[Minchev et al. 2012a]{minchev12a} and discussion therein).

\subsection{The chemistry}
\label{sec:chem}

The chemical evolution model we use here is for the \emph{thin disc only}. The idea behind this is to test, once radial mixing is taken into account, if we can explain the observations of both thin and thick discs without the need of invoking a discrete thick-disc component. A detailed description of the chemical model is given in \cite[Minchev et al. (2013)]{minchev12c} and here we show its properties in Fig.~\ref{fig2}.

\subsection{Fusing chemistry and dynamics}

To combine the dynamics with the chemistry, we start by dividing the simulated disc into 300-pc radial bins in the range $0<r<16$~kpc. At each time output we randomly select newly born stars, by matching the SFH corresponding to our chemical evolution model at each radial bin. Since our goal is to satisfy the SFH expected for the chemical model (close, but not identical to that of the simulation), at some radial bins and times we may not have enough newly born stars in the simulation. In such a case we randomly select stars currently on the coldest orbits (i.e., with kinematics close to those of newly born stars). This fraction is about 10\% at the final time. The particles selected at each time output are assigned the corresponding chemistry for that particular radius and time. This is done recursively at each time-step of 37.5~Myr. For the current work we extract about 40\% of the $\approx2.5x10^6$ stellar particles comprising our dynamical disc at the final time in the range $|z|<4$~kpc, $r<16$~kpc. We have verified that decreasing or increasing this extracted sample by a factor two has negligible effects on the results we present in this work.

In the just described manner, we follow the self-consistent disc evolution for 11.2~Gyr, selecting a tracer population possessing known SFH and chemistry enrichment. This bypasses all the problems encountered by previous chemodynamical models based on N-body simulations. It also offers a way to easily test the impact of different chemical prescriptions on the chemodynamical results. 

\subsection{Local chemodynamical relations}

Fig.~\ref{fig3} shows how radial migration affects the simulated solar neighborhood. A detailed discussion can be found in \cite[Minchev et al. (2013)]{minchev12c}. Below we concentrate on the comparison of the model to observations.

\begin{figure}
\includegraphics[width=5.3in]{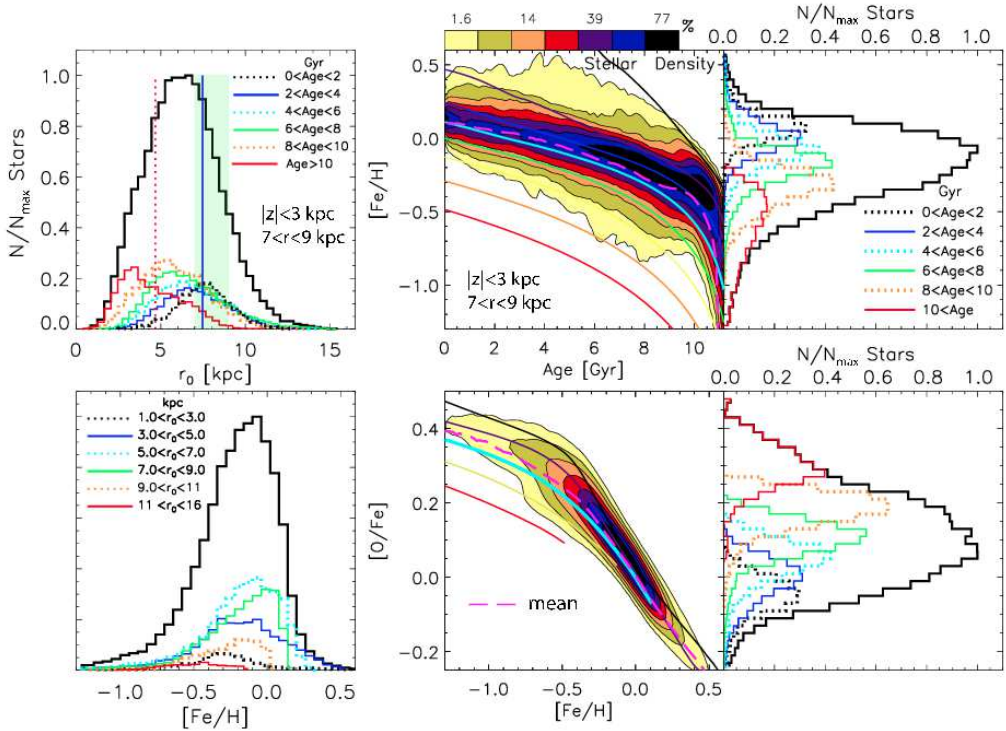}
\caption{
{\bf Top left:} Birth radii of stars ending up in the ``solar" radius (green shaded strip) at the final simulation time. The solid black curve plots the total $r_0$-distribution, while the color-coded curves show the distributions of stars in six different age groups, as indicated. The dotted-red and solid-blue vertical lines indicate the positions of the bar's CR and OLR at the final simulation time. A large fraction of old stars comes from the inner disc, including from inside the CR. {\bf Bottom left:} [Fe/H] distributions for the sample shown in the top panel, binned by birth radius in six groups, as indicated. The total distribution is shown by the solid black curve. The importance of the bar's CR is seen in the large fraction of stars with $3<r_0<5$~kpc (blue line) giving rise to the metal-rich tail of the distribution. 
{\bf Top middle:} Stellar density contours of the Age-[Fe/H] relation after fusing with dynamics, for the ``solar" radius ($7<r<9$~kpc, as in left column). The overlaid curves indicate the input chemistry for certain radii, indicating where stars were born. The pink dashed curve plots the mean, seen to follow closely the input $r_0=8$~kpc line (cyan). 
{\bf Bottom middle:} Same as top but for the [Fe/H]-[O/Fe] relation. No bimodality is found for this unbiased model sample.
{\bf Top right:} Metallicity distributions for different age bins. {\bf Bottom right:} Same as above but for the [Fe/H]-[O/Fe] relation.
}
\label{fig3}      
\end{figure}

\begin{figure}
\centerline{\includegraphics[width=5.3in]{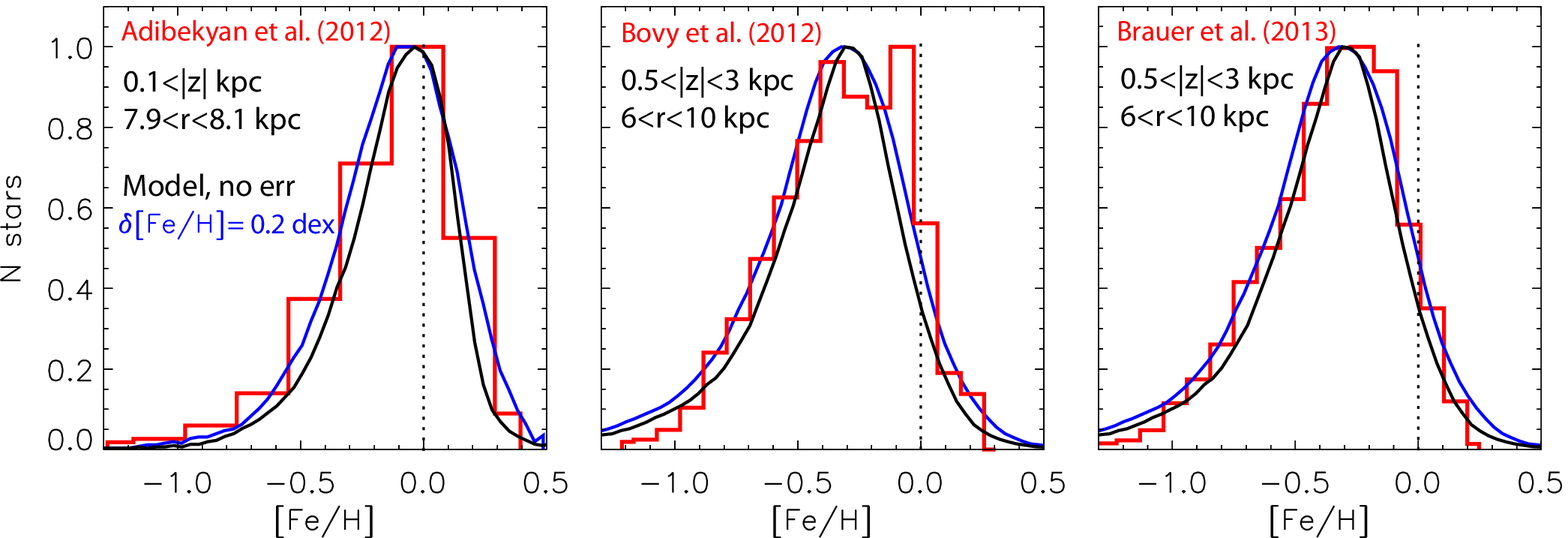}}
\caption{
Comparison of our model metallicity distribution function (MDF) to recent observations. The red histograms shows data from \cite[Adibekyan et al. (2012)]{adibekyan12} (left), \cite[Bovy et al. (2012a)]{bovy12a} (middle), and Brauer et al. (2013) (right). In each panel the black and blue curves plot the model data with no error in [Fe/H] and convolved error of $\delta$[Fe/H]=0.2~dex, respectively. We have selected a very local sample to compare to the high-resolution nearby sample by \cite[Adibekyan et al. (2012)]{adibekyan12} (as indicated) and a range in $z$ and $r$ to approximately match the SEGUE data studied by the other two works. A shift in the peak from [Fe/H]$\sim$0 to [Fe/H]$\sim-0.3$ dex for both observed and simulated data is seen when considering a larger sample depth. 
}
\label{fig4}      
\end{figure}

\section{Comparison to recent observational results}

To be able to compare our results to different MW surveys we need to constrain spatially our model sample. For example the Geneva-Copenhagen Survey (GCS) and high-resolution stellar samples are confined to within 100~pc from the Sun, while the SEGUE and RAVE samples explore much larger regions, but miss the local stars. 

\subsection{Changes in the MDF for samples at different distances from the disc plane}
\label{sec:segue}

In the right panel of Fig.~\ref{fig4} we compare our model MDF to recent observations. We consider the high-resolution sample by \cite[Adibekyan et al. (2012)]{adibekyan12} (histogram extracted from Fig.~16 by \cite[Rix and Bovy 2013]{rix13}), the mass-corrected SEGUE G-dwarf sample by \cite[Bovy et al. (2012a)]{bovy12a} (MDF extracted from their Fig.~1, right panel). The third data set we consider comes from the latest, Data Release (DR) 9, SEGUE G-dwarf sample, which is also mass-corrected in the same way as \cite[Bovy et al. (2012a)]{bovy12a}. The main differences from the data used by \cite[Bovy et al. (2012a)]{bovy12a} are a different distance estimation (spectroscopic instead of photometric), a selection of G-dwarfs based on DR9, instead of on DR7 photometry, and a signal-to-noise $S/N>20$ instead of $S/N>15$. Further details can be found in Brauer et al. (2013, in preparation). The exact cut in $r$ and $z$ as in the simulation is applied to the latter sample only, since the other two data sets are not available to us.

The red histogram in each panel of Fig.~\ref{fig4} shows different observational data as indicated; the additional curves plot the model data with no error in [Fe/H] (black) and convolved errors of $\delta$[Fe/H]=$\pm0.2$~dex (blue), drawn from a uniform distribution. We have selected a very local sample to compare to the high-resolution nearby data by \cite{adibekyan12} and a range in $z$ and $r$ to approximately match the SEGUE data studied by the other two works. A remarkable match is found between our model predictions and the data. The expected broadening in the model distributions is seen, with the inclusion of error in [Fe/H].

\subsection{Is the bimodality in the [Fe/H]-[O/Fe] plane due to selection effects?}
\label{sec:sel}

Starting with the assumption of two distinct entities -- the thin and thick discs -- many surveys select stars according to certain kinematical criteria (e.g., \cite[Bensby et al. 2003, Reddy et al. 2006]{bensby03, reddy06}). For example, \cite[Bensby et al. (2003)]{bensby03} defined a (now widely used) method for preferentially selecting thin- and thick-disc stars with a probability function purely based on kinematics. 

We now employ the same technique as \cite[Bensby et al. (2003)]{bensby03} and extract a thin- and thick-disc selections from our initially unbiased sample (Fig.~\ref{fig3}, bottom middle panel). In order to obtain a similar number of thin- and thick-disc stars, as done in most surveys, we randomly down-sample the thin-disc selection. To mach the spacial distribution of high-resolution samples, usually confined to small distances from the Sun, we impose the constraints $7.8<r<8.2$~kpc, $|z|<0.2$~kpc. Our results do not change to any significant degree when this sample depth is decreased or increased by a factor of two.

\begin{figure}
\centerline{\includegraphics[width=5.3in]{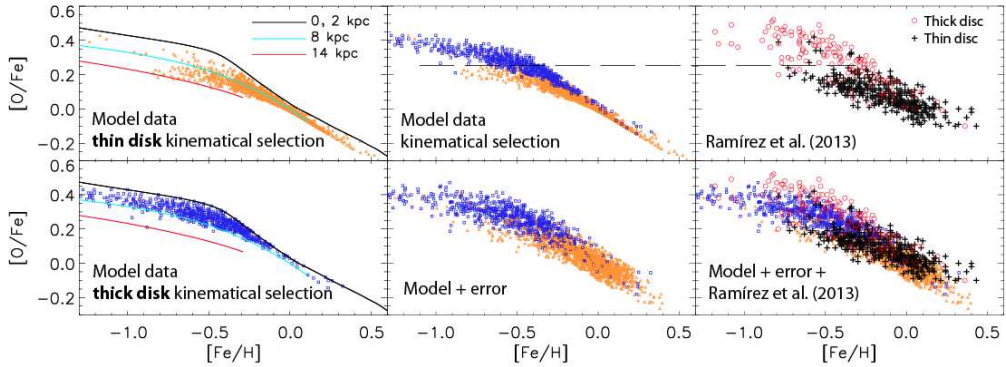}}
\caption{
Selection effects can result in a bimodality in the [Fe/H]-[O/Fe] plane. {\bf Left column:} We have applied the selection criteria used by \cite[Bensby et al. (2003)]{bensby03} to extract thin- and thick-disc samples from our model, indicated by orange triangles and blue squares, respectively. {\bf Middle column:} Overlaid thin- and thick-disc model samples, which are shown separately in the left column. The bottom panel shows the same model data but with convolved errors of $\delta$[Fe/H]=$\pm0.1$ and $\delta$[O/Fe]=$\pm0.05$~dex. {\bf Right column:} Top panel plots the high-resolution data from \cite[Ram{\'{\i}}rez]{ramirez13} (their Fig.~11, b). In the bottom panel we overlay the observed and simulated data samples, presenting a remarkable match. The black dashed horizontal line shows the location of the gap coincidental for both the model and data.
}
\label{fig5}      
\end{figure}

The resulting thin- and thick-disc distributions of stars in the [Fe/H]-[O/Fe] plane are presented, respectively, in the top and bottom left panels of Fig.~\ref{fig5}. To get an idea of where stars originate, we have overlaid the input chemistry curves for stars born at $r_0\leqslant2$~kpc (black), at $r_0=8$~kpc (or in situ, cyan), and at $r_0=14$~kpc (red). The thick-disc selection appears to originate almost exclusively inside the solar circle (sample is confined between the cyan and black curves). On the other hand, the thin-disc sample follows the local curve for [Fe/H]~$>0.1$, covers the region between the inner and outer disc (black and red curves) for $-0.4<$~[Fe/H]~$<0.1$. Thin-disc stars with [Fe/H]~$<-0.4$, referred to as ``the metal-poor thin disc" (e.g., \cite[Haywood et al. 2013]{haywood13}), appear to originate mostly from outside the solar radius in agreement with the conclusions by \cite[Haywood et al. (2013)]{haywood13}.

The middle, top panel of Fig.~\ref{fig5} displays the overlaid thin- and thick-disc selections, which were showed separately in the left column. A relatively clear bimodality is found at [O/Fe]~$\approx0.2$~dex, similarly to what is seen in observational studies. The thick-disc selection follows the upper boundary of the distribution throughout most of the [Fe/H] extent due to the fact that it originates in the inner disc. In the bottom, middle panel we have convolved errors within our model data of $\delta$[Fe/H]=$\pm0.1$ and $\delta$[O/Fe]=$\pm0.05$ dex, drawing from a uniform distribution.

In the right, top panel of Fig.~\ref{fig5} we plot the high-resolution data from \cite{ramirez13} (same as their Fig.~11, b). The gap separating the thin- and thick-disc selections occurs at a very similar location to what is found for the model on the left, as indicated by the black dashed horizontal line. In the bottom right panel we overlay the observed and simulated data samples, presenting a remarkable match. An overall offset within the error ($0.05-0.1$~dex) toward negative [O/Fe]-values is found for the model.

\begin{figure}
\centerline{\includegraphics[width=4.3in]{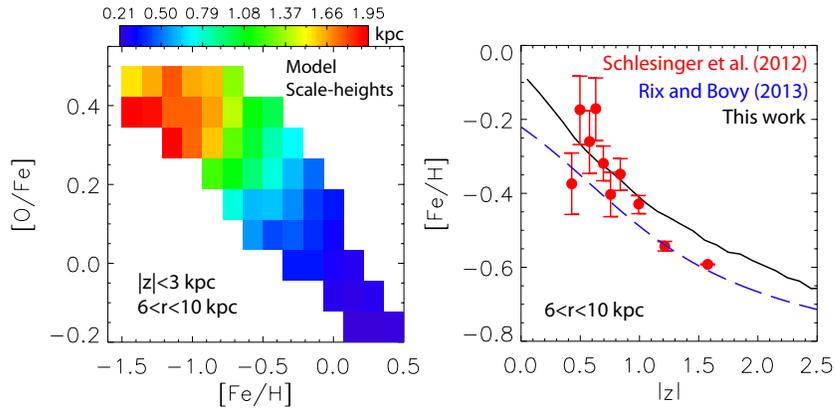}}
\caption{
{\bf Left:} Model vertical scale-heights for different mono-abundance subpopulations as a function of position in the [Fe/H]-[O/Fe] plane. Very good fits are obtained by using single exponentials, excluding the 200~pc closest to the disc plane. A radial range of $6<r<10$~kpc similar to the SEGUE data is used. This figure can be directly compared to Fig.~4 by \cite[Bovy et al. (2012b)]{bovy12b}. Our model predicts scale-heights of up to $\sim1.9$~kpc for the oldest, most metal-poor samples; this should be compared to a maximum of $\sim1$~kpc inferred from Bovy et al.'s mass corrected SEGUE G-dwarf data.
{\bf Right:} Mean metallicity as a function of distance from the disc plane, $|z|$, in the range $6<r<10$~kpc. Red filled circles and error bars show SEGUE data from \cite[Schlesinger et al. (2012)]{schlesinger12}, while blue dashed line is Bovy et al.'s model. The solid black curve is our model. A very good agreement between the two models in found for the metallicity variation with $|z|$, including the flattening at $|z|\gtrsim1.5$~kpc. This can be associated with the predominance of old/high-[O/Fe] stars at large distances from the plane.
}
\label{fig6}      
\end{figure}

\subsection{Disc scale-heights of mono-abundance subpopulations}

Using SEGUE G-dwarf data, \cite[Bovy et al. (2012b)]{bovy12b} (see also Bovy et al., this volume) showed that the vertical density of stars grouped in narrow bins constrained in both [O/Fe] and [Fe/H] (mono-abundance subpopulations) can be well approximated by single exponential disc models. 

To see whether we can reproduce these results, in the left panel of Fig.~\ref{fig6} we present the vertical scale-heights of different mono-abundance subpopulations, resulting from our model, as a function of their position in the [Fe/H]-[O/Fe] plane. We have convolved errors in our model data of 0.1 and 0.05 dex for [Fe/H] and [O/H], respectively. A range in galactic radius of $6<r<10$~kpc, similar to the SEGUE data, is used. Very good fits are obtained by using single exponentials, excluding the 200~pc closest to the disc plane. 

This figure can be directly compared to Fig.~4 by \cite[Bovy et al. (2012b)]{bovy12b}, bottom left panel. The general trends found in the SEGUE G-dwarf data are readily reproduced by our model: (i) for a given metallicity bin, scale-heights decrease when moving from low- to high-[O/Fe] populations and (ii) for a given [O/Fe] bin, scale-heights increase with increasing metallicity. 

Note that a reversal is found for trend (i) above, just as in the data (Fig.~4, \cite[Bovy et al. 2012b]{bovy12b}): for the four most metal-poor bins ([Fe/H]~$<-0.7$~dex) scale-heights {\it decrease} with increasing [O/Fe]. This may be a potentially very important observation, that can be linked to the MW merger history as recently described by \cite[Minchev et al. (2014)]{minchev14}.

\subsection{The vertical metallicity gradient}

We would now like to compare our model to the recent works by \cite[Schlesinger et al. (2012)]{schlesinger12} and \cite[Rix and Bovy (2013)]{rix13}, where SEGUE G-dwarf data were utilized. We have extracted the data points from Fig.~15 by \cite[Rix and Bovy (2013)]{rix13}, where a comparison was made between the vertical metallicity gradient estimated by \cite[Schlesinger et al. (2012)]{schlesinger12} and the one implied by Bovy et al.'s model.

The right panel of Fig.~\ref{fig6} presents these data along with our model prediction, as indicated by different colors. We find $\sim0.05-1$~dex higher metallicity than Bovy et al.'s model systematically as a function of $|z|$, appearing to match the data better at $|z|\lesssim1$~kpc. We note that both models and the data are normalized using the solar abundance values by \cite[Asplund et al. (2005)]{asplund05}. It should be kept in mind that while the errors in [Fe/H] are plotted for each data point, distance uncertainties are not shown; however they should be increasing with distance from the disc plane. 

A very good agreement between the two models is found for the metallicity variation with $|z|$, including the flattening at $|z|\gtrsim1.5$~kpc. \cite[Minchev et al. (2013)]{minchev12c} associated this with the predominance of old/high-[O/Fe] stars at large distances from the plane.

\section{Discussion}

Despite the recent advances in the general field of galaxy formation and evolution, there are currently no self-consistent simulations that have the level of chemical implementation required for making detailed predictions for the number of ongoing and planned MW observational campaigns. Even in high-resolution N-body experiments, one particle represents $\sim10^{4-5}$ solar masses. Hence, a number of approximations are necessary in order to compute the chemical enrichment self-consistently in a simulation, leading to the several problems briefly discussed in Sec.~\ref{sec:cos}. Here, instead, we have assumed that each particle is one star\footnote{Dynamically, this is a good assumption, since the stellar dynamics is collisionless.} and have implemented the exact SFH and chemical enrichment from a typical chemical evolution model into a state-of-the-art simulation of the formation of a galactic disc. We note that this is the first time that a chemodynamical model has the extra constraint of defining a realistic solar vicinity also in terms of dynamics, as detailed in Sec.~\ref{sec:model}.

According to our model, the MW ``thick disc" emerges naturally from (i) stars born with high velocity dispersions at high redshift, (ii) stars migrating from the inner disc very early on due to strong merger activity, and (iii) further radial migration driven by the bar and spirals at later times. Importantly, a significant fraction of old stars with thick-disc characteristics could have been born near and outside the solar radius (Fig.~15 by \cite[Minchev et al. 2013]{minchev12c}).

\cite[Minchev et al. (2013)]{minchev12c} showed that for a simulation lacking early mergers the velocity dispersion of the oldest disc stars in the simulated solar neighborhood is underestimated by a factor of $\sim2$, while the model presented here is compatible with the observations (e.g., the vertical age-velocity relation achieves a maximum at $\sim50$~km/s for the oldest stars, see Fig.~6 by \cite[Minchev et al. 2013]{minchev12c}; Smith et al., this volume). It is, therefore, tempting to conclude that the highest velocity dispersion stars observed in the solar neighborhood are a clear indication of an early-on, merger-dominated epoch, or equivalently, scenarios involving stars born with high velocity dispersions. Our results suggest that the MW thick disc cannot be explained by a merger-free disc evolution as proposed previously. This is related to the inability of internally driven radial migration to thicken galactic discs due to the conservation of vertical action (\cite[Minchev et al. 2012b]{minchev12b}). However, an observational test involving both chemical and kinematic information must be devised to ascertain the effect of early mergers in the MW.

The results of this work present an improvement over previous models as we use a state-of-the-art simulation of a disc formation to extract self-consistent dynamics and fuse with a chemical model tailored for the MW. Future work should consider implementing gas flows, Galactic winds, and Galactic fountains (Fraternali et al., this volume) into the model to asses the importance of these processes. A suite of different chemical evolution models assigned to different disc dynamics (from different simulations) should be investigated to make progress in the field of Galactic Archeology. 

The method presented here is not only applicable to the MW but is also potentially very useful for extragalactic surveys. The same approach can be used for any other galaxy for which both kinematic and chemical information is available. A large such sample is currently becoming available from the ongoing CALIFA survey (\cite[Sanchez et al. 2012]{sanchez12}). Given the SFH and morphology of each galaxy, a chemodynamical evolution model tailored to each individual object can be developed, thus providing a state-of-the-art database for exploring the formation and evolution of galactic discs.

\begin{discussion}

\discuss{Kobayashi}{In your model, what are the physical reasons for the metallicity gradients not to evolve as a function of time? Did you change the SF timescale as a function of radius?}

\discuss{Minchev}{In fact the gas metallicity gradient varies both as a function of time and radius. The chemical model we use is similar to the thin-disc model by Chiappini (2009), where the disc grows inside-out as in the cosmological re-simulation we use. A detailed description can be found in Minchev et al. (2013).}

\end{discussion}


\begin{thebibliography}{}

\bibitem[\protect\astroncite{{Abadi} et~al.}{2003}]{abadi03}
{Abadi}, M.~G., {Navarro}, J.~F., {Steinmetz}, M., and {Eke}, V.~R. 2003,
\textit{ApJ}, 597, 21

\bibitem[Adibekyan et al.(2012)]{adibekyan12} 
Adibekyan, V.~Z., Sousa, S.~G., Santos, N.~C., et al.\ 2012, 
\textit{A\&A}, 545, A32

\bibitem[\protect\astroncite{{Agertz} et~al.}{2011}]{agertz11}
{Agertz}, O., {Teyssier}, R., and {Moore}, B. 2011,
\textit{MNRAS}, 410, 1391

\bibitem[Asplund et al.(2005)]{asplund05} 
Asplund, M., Grevesse, N., \& Sauval, A.~J.\ 2005,
Cosmic Abundances as Records of Stellar Evolution and Nucleosynthesis, 336, 25

\bibitem[\protect\astroncite{{Bensby} et~al.}{2003}]{bensby03}
{Bensby}, T., {Feltzing}, S., and {Lundstr{\"o}m}, I. 2003,
\textit{A\&A}, 410, 527

\bibitem[\protect\astroncite{{Bovy} et~al.}{2012a}]{bovy12a}
{Bovy}, J., {Rix}, H.-W., and {Hogg}, D.~W. 2012a,
\textit{ApJ}, 751, 131

\bibitem[\protect\astroncite{{Bovy} et~al.}{2012b}]{bovy12b}
{Bovy}, J., {Rix}, H.-W., {Liu}, C., {Hogg}, D.~W., {Beers}, T.~C., and {Lee},
Y.~S. 2012b, 
\textit{ApJ}, 753, 148

\bibitem[Brook et al.(2012)]{brook12} Brook, C.~B., Stinson, 
G.~S., Gibson, B.~K., et al.\ 2012, 
\textit{MNRAS}, 426, 690 

\bibitem[Calura et al.(2012)]{calura12} 
Calura, F., Gibson, B.~K., Michel-Dansac, L., et al.\ 2012, 
\textit{MNRAS}, 427, 1401

\bibitem[Zhao et al.(2006)]{zhao06} 
Zhao, G., Chen, Y.-Q., Shi, J.-R., et al.\ 2006, 
\textit{CJAA}, 6, 265

\bibitem[\protect\astroncite{{de Jong} et~al.}{2012}]{dejong12}
{de Jong}, R.~S., {Bellido-Tirado}, O., and {Chiappini}, C. et~al. 2012, Procspie, 8446

\bibitem[\protect\astroncite{{Dehnen}}{2000}]{dehnen00}
{Dehnen}, W.: 2000,
\textit{AJ}, 119, 800

\bibitem[\protect\astroncite{{Few} et~al.}{2012}]{few12}
{Few}, C.~G., {Courty}, S., {Gibson}, B.~K., et al. 2012,
\textit{MNRAS}, 424, L11

\bibitem[\protect\astroncite{Freeman 2010}]{freeman10}
Freeman, K. C. 2010, Galaxies and their Masks, 319

\bibitem[\protect\astroncite{Gilmore et al. 2012}]{gilmore12}
Gilmore, G., Randich, S., Asplund, M. et al. 2012, The Messenger 147, 25 

\bibitem[\protect\astroncite{{Guedes} et~al.}{2011}]{guedes11}
{Guedes}, J., {Callegari}, S., {Madau}, P., and {Mayer}, L. 2011,
\textit{ApJ}, 742, 76

\bibitem[Haywood et al.(2013)]{haywood13} 
Haywood, M., Di Matteo, P., Lehnert, M., Katz, D., \& Gomez, A.\ 2013, 
ArXiv

\bibitem[\protect\astroncite{{Kawata} and {Gibson}}{2003}]{kawata03}
{Kawata}, D. and {Gibson}, B.~K. 2003,
\textit{MNRAS}, 340, 908

\bibitem[\protect\astroncite{{Kobayashi} et~al.}{2007}]{kobayashi07}
{Kobayashi}, C., {Springel}, V., and {White}, S.~D.~M.: 2007,
\textit{MNRAS}, 376, 1465

\bibitem[\protect\astroncite{{Kobayashi} and {Nakasato}}{2011}]{kobayashi11}
{Kobayashi}, C. and {Nakasato}, N. 2011,
\textit{ApJ}, 729, 16

\bibitem[\protect\astroncite{{Majewski} et~al.}{2010}]{majewski10}
{Majewski}, S.~R., {Wilson}, J.~C., {Hearty}, F., et al. 2010,
in K. {Cunha}, M. {Spite}, and B. {Barbuy} (eds.), {\em IAU
  Symposium}, Vol. 265 of {\em IAU Symposium}, pp 480--481
 
\bibitem[Martig et al.(2012)]{martig12} 
Martig, M., Bournaud, F., Croton, D.~J., Dekel, A., \& Teyssier, R.\ 2012, 
\textit{ApJ}, 756, 26 

\bibitem[\protect\astroncite{{Martig} et~al.}{2009}]{martig09}
{Martig}, M., {Bournaud}, F., {Teyssier}, R., and {Dekel}, A.: 2009,
\textit{ApJ}, 707, 250

\bibitem[\protect\astroncite{{Minchev} et~al.}{2007}]{mnq07}
{Minchev}, I., {Nordhaus}, J., and {Quillen}, A.~C.: 2007,
\textit{ApJ} (Letters), 664, L31

\bibitem[Minchev et al.(2012a)]{minchev12a} 
Minchev, I., Famaey, B., Quillen, A.~C., et al.\ 2012a, 
\textit{A\&A}, 548, A126

\bibitem[Minchev et al.(2012b)]{minchev12b} 
Minchev, I., Famaey, B., Quillen, A.~C., et al.\ 2012b, 
\textit{A\&A}, 548, 127

\bibitem[Minchev et al.(2013)]{minchev12c} 
Minchev, I., Chiappini, C., \& Martig, M.\ 2013, 
\textit{A\&A}, 558, A9  

\bibitem[Minchev et al.(2014)]{2014ApJ...781L..20M} Minchev, I., Chiappini, 
C., Martig, M., et al.\ 2014, 
\textit{ApJ} (Letters), 781, L20

\bibitem[\protect\astroncite{{Navarro} and {Benz}}{1991}]{navarro91}
{Navarro}, J.~F. and {Benz}, W. 1991,
\textit{ApJ}, 380, 320

\bibitem[\protect\astroncite{{Peng} et~al.}{2002}]{peng02}
{Peng}, C.~Y., {Ho}, L.~C., {Impey}, C.~D., and {Rix}, H.-W.: 2002,
\textit{AJ}, 124, 266

\bibitem[\protect\astroncite{Perryman et al. 2001}]{perryman01}
Perryman, M. A. C., de Boer, K. S., Gilmore, G., et al. 2001, 
\textit{A\&A}, 369, 339 


\bibitem[\protect\astroncite{{Raiteri} et~al.}{1996}]{raiteri96}
{Raiteri}, C.~M., {Villata}, M., and {Navarro}, J.~F.: 1996,
\textit{A\&A}, 315, 105

\bibitem[Ram{\'{\i}}rez et al.(2013)]{ramirez13} 
Ram{\'{\i}}rez, I., Allende Prieto, C., \& Lambert, D.~L.\ 2013,
\textit{ApJ}, 764, 78

\bibitem[\protect\astroncite{{Reddy} et~al.}{2006}]{reddy06}
{Reddy}, B.~E., {Lambert}, D.~L., and {Allende Prieto}, C. 2006,
\textit{MNRAS}, 367, 1329

\bibitem[Rix \& Bovy(2013)]{rix13} 
Rix, H.-W., \& Bovy, J.\ 2013, 
\textit{A\&ARv}, 21, 61

\bibitem[S{\'a}nchez et al.(2012)]{sanchez12} 
S{\'a}nchez, S.~F., Kennicutt, R.~C., Gil de Paz, A., et al.\ 2012, 
\textit{A\&A}, 538, A8

\bibitem[\protect\astroncite{{Scannapieco} et~al.}{2005}]{scannapieco05}
{Scannapieco}, C., {Tissera}, P.~B., {White}, S.~D.~M., and {Springel}, V. 2005, 
\textit{MNRAS}, 364, 552

\bibitem[\protect\astroncite{{Steinmetz} et~al.}{2006}]{steinmetz06}
{Steinmetz, M., Zwitter, T., Siebert}, et~al. 2006,
\textit{AJ}, 132, 1645

\bibitem[\protect\astroncite{{Tissera} et~al.}{2012}]{tissera12}
{Tissera}, P.~B., {White}, S.~D.~M., and {Scannapieco}, C.: 2012,
\textit{MNRAS}, 420, 255

\bibitem[\protect\astroncite{{Wiersma} et~al.}{2009}]{wiersma09}
{Wiersma}, R.~P.~C., {Schaye}, J., {Theuns}, et al. 2009, 
\textit{MNRAS}, 399, 574

\bibitem[\protect\astroncite{Yanny, B. et al.}{2009}]{yanny09}
{Yanny, B., Rockosi, C., N. H. J.} et al. 2009, 
\textit{AJ}, 137, 4377 

\end{thebibliography}
\end{document}